\documentclass[10pt, conference]{IEEEtran}
\IEEEoverridecommandlockouts
\usepackage{cite}
\usepackage{verbatim}
\usepackage{url}
\usepackage{grffile}
\usepackage{graphicx}
\usepackage{subcaption}
\usepackage{amsmath,amssymb,amsfonts}
\usepackage[detect-all]{siunitx}
\usepackage[hidelinks]{hyperref}
\usepackage{cleveref}
\Crefname{section}{Sec.}{Secs.}
\usepackage{tikz}
\usepackage{algorithmic}
\usepackage{epstopdf}
\usepackage[utf8]{inputenc}

\def\BibTeX{{\rm B\kern-.05em{\sc i\kern-.025em b}\kern-.08em
    T\kern-.1667em\lower.7ex\hbox{E}\kern-.125emX}}

\newcommand\copyrighttext{%
  \footnotesize \textcopyright 2024 IEEE. Personal use of this material is permitted.
  Permission from IEEE must be obtained for all other uses, in any current or future 
  media, including reprinting/republishing this material for advertising or promotional purposes, creating new collective works, for resale or redistribution to servers or lists, or reuse of any copyrighted component of this work in other works.}
\newcommand\copyrightnotice{%
\begin{tikzpicture}[remember picture,overlay]
\node[anchor=south,yshift=10pt] at (current page.south) {\fbox{\parbox{\dimexpr\textwidth-\fboxsep-\fboxrule\relax}{\copyrighttext}}};
\end{tikzpicture}%
}    
    
\usepackage[]{footmisc}

\usepackage{booktabs}

\begin{document}
\bstctlcite{IEEEexample:BSTcontrol}

\title{RACH-less Handover with Early Timing Advance Acquisition for Outage Reduction  \vspace{-0.2\baselineskip}\\
}

\author{
    \IEEEauthorblockN{
        Subhyal Bin Iqbal\IEEEauthorrefmark{1}\IEEEauthorrefmark{2}, Umur Karabulut\IEEEauthorrefmark{1}, Ahmad Awada\IEEEauthorrefmark{1}, Philipp Schulz\IEEEauthorrefmark{2} and Gerhard P. Fettweis\IEEEauthorrefmark{2}  
    }

    \IEEEauthorblockA{\IEEEauthorrefmark{1} Nokia Standardization and Research Lab, Munich, Germany}
    \IEEEauthorblockA{\IEEEauthorrefmark{2} Vodafone Chair for Mobile Communications Systems, Technische Universität Dresden, Germany}
\vspace{-2.3\baselineskip}
}
\maketitle

% *** IEEE Copyright notice with TikZ ***
% 
\copyrightnotice

\begin{abstract}
For fifth-generation (5G) and 5G-Advanced networks, outage reduction within the context of reliability is a key objective since outage denotes the time period when a user equipment (UE) cannot communicate with the network. Earlier studies have shown that in the experimental high mobility scenario considered, outage is dominated by the interruption time that stems from the random access channel (RACH)-based handover process from the serving cell to the target cell. A handover by itself is a necessary mobility process to prevent mobility failures and their associated outage. This paper proposes a RACH-less handover signaling scheme for the 3rd Generation Partnership Project (3GPP) conditional handover (CHO) mechanism. The proposed scheme exploits the decoupling between the CHO preparation and execution phases to establish initial synchronization between the UE and the target cell through an early acquisition of the timing advance. This significantly curtails the RACH process and therefore the handover interruption time. Results based on a system-level simulation-based mobility study have shown that the proposed scheme significantly reduces the outage and its constituent handover interruption time relatively by 18.7\% and 43.2\%, respectively.
\end{abstract} 


\section{Introduction} \label{Section1}
 
One of the major objectives of fifth-generation (5G) and 5G-Advanced networks\cite{5GAdvanced} is outage reduction to improve the overall network reliability, specifically in frequency range 2 (FR2) \cite{38331}. This is especially needed for ultra-reliable low-latency communications (URLLC). In simple terms, an outage can be defined as the time duration during which the user equipment (UE) cannot receive data from the network \cite{36881}. One component of outage is the service or handover interruption time, which has been defined in 3rd Generation Partnership Project (3GPP) terminology \cite{HOInterruptionTimeReductionTechRep} as the time duration during a handover when the UE exchanges no data with the serving cell anymore and none with the target cell yet. In our earlier study \cite{Rxbeamformingpaper} we have already shown through simulation-based investigations that in a 5G-beamformed network operating in FR2 with UE speeds of 60 km/h, the most significant outage contribution stems from handover interruption time. The other components are mobility failures and signal-to-noise-plus-interference ratio (SINR) degradation. Hence, reducing the handover interruption time is critical in reducing the overall outage in the network and improving the network reliability.

The handover interruption time is primarily governed by the random access channel (RACH) procedure \cite{36300}, whereby the UE obtains uplink synchronization with the target cell while executing a handover. Recently, in \textit{3GPP Release 15} \cite{36300} RACH-less handover has been proposed whereby the RACH procedure can be simplified or avoided altogether. In order for a RACH-less handover to bear fruition, the UE has to acquire the uplink timing advance (TA) and the associated uplink (UL) transmission grant \cite{36300} of the target cell, amongst others. This has to be done before the execution of the actual handover itself. While 3GPP has proposed RACH-less handover, the exact methodology for an early TA acquisition has not been specified and has been left open to implementation, dependent on the underlying handover mechanism.

The authors in \cite{NokiaRACHlesspaper} model RACH-less handover for the legacy baseline handover (BHO) mechanism \cite{38300} in a time-synchronous network. In such networks, the UE can derive the TA of the target cell based on the TA of the source cell and the time difference in the signals received from the source and target cells. Thus, the RACH procedure can largely be avoided. On the other hand, the authors in \cite{UniveristyRACHlesspaper} also model RACH-less handover for BHO but for an asynchronous network, using a similar approach for TA acquisition together with additional time alignment. The authors in \cite{UniveristyRACHlesspaper2} use a similar approach to \cite{UniveristyRACHlesspaper} for the BHO mechanism, but additionally check for the accuracy of calculated UE-calculated TA value, upon which an additional TA acquisition is performed. In doing so, all approaches slightly differ from the 3GPP standard since it is the target cell's role to estimate the TA based on the RACH preamble transmission \cite{JedrzejRACHless}.

Conditional handover (CHO) is a state-of-the-art handover mechanism introduced in \textit{3GPP Release 16} \cite{CHOTechRep} as an alternative to the legacy baseline handover to improve mobility performance in 5G mobile networks. This is achieved by decoupling the actual handover preparation from the handover execution by introducing a conditional procedure. This paper proposes an elaborate and novel signaling scheme whereby the UE acquires the TA of the target cell before the handover execution by exploiting the split between the CHO preparation and execution conditions. In doing so, this paper not only conforms to the standard but also models RACH-less handover for the more advanced CHO mechanism. Thereafter, the mobility performance is analyzed in a system-level simulation for a 5G-beamformed network, where the mobile UE is modeled to correspond to the multi-panel UE (MPUE) architecture \cite{SubhyalMPUEpaper}. It is pertinent to mention that MPUEs are an essential part of 5G-Advanced and future 3GPP releases in the context of FR2 deployments. To the best of the authors' knowledge, a system-level mobility performance analysis of RACH-less handover based on early TA acquisition for CHO has not been investigated before, providing a basis for a major contribution to this paper. The results are examined to study the impact of the proposed handover scheme in reducing the outage in the network, together with an analysis of the signaling overhead of the proposed scheme. 

 %A conference paper. Citing \cite{fastCHOpaper} and the book \cite{dahlman5Gbook}, among others.

%%%%%%%%%%%%%%%%%%%%%%

\section{Network Model} \label{Section2}
%In this section, the inter-cell and intra-cell mobility that form part of the handover and beam management procedures, respectively, are discussed along with the SINR model. 

This section discusses the inter-cell mobility for the handover procedure, the intra-cell mobility for the beam management procedures, and the SINR model.

\subsection{Inter-cell Mobility} \label{Subection2.1}
Inter-cell mobility refers to handovers between different cells in the network. For a successful handover from the serving cell $c_0$ to the target cell $c^{\prime}$, a pre-requisite is the filtering of physical layer reference signal received power (RSRP) measurements to mitigate the effect of rapid channel impairments. When a 5G beamformed network is considered, each UE is assumed to be capable of measuring the raw RSRP values $P_{c,b}^\textrm{RSRP}(n)$ at a discrete time instant $n$ from each Tx beam $b \in B$ of cell $c \in C$, using the synchronization signal block (SSB) bursts that are periodically transmitted by the base station (BS). The separation between the time instants is denoted by $\Delta t$ ms. At the UE side, layer 1 (L1) and L3 filtering are then sequentially applied to the raw RSRPs to counter the effects of fast fading and measurement errors. The output of the L1 filter is the L1 RSRP $P_{c,b}^\textrm{L1}(m)$. Here $m = n\omega$, with $\omega$ being the L1 measurement period (aligned with the SSB periodicity) normalized by the time step duration $\Delta t$. At the end of the L3 filtering, the UE has determined the L3 cell quality for the serving and neighboring cells. A more detailed explanation of the L1 and L3 filtering procedures can be found in \cite[Sec. II.A]{SubhyalMPUEpaper}. 

L3 cell quality $P_{c}^\textrm{L3}(m)$ is an important indicator of the average downlink signal strength for a link between a UE and cell~$c$ and is therefore used in the handover process. As mentioned in \Cref{Section1}, CHO decouples handover preparation and execution by preparing the handover early. Still, the actual handover execution only occurs when the radio link is sufficient. In the \textit{CHO preparation phase}, the UE initiates the preparation of the target cell $c^{\prime}$. As such, the CHO preparation condition is monitored by the UE and is defined as
\begin{equation}
\label{Eq1}
     P_{c_0}^\textrm{L3}(m)  < P_{c^{\prime}}^\textrm{L3}(m) + o^\textrm{prep}_{c_0,c^{\prime}} \ \text{for} \  m_\textrm{0} - T_\textrm{prep} < m < m_\textrm{0},
\end{equation} 
where $o^\textrm{prep}_{c_0,c^{\prime}}$ is defined as the CHO preparation offset between cell $c_0$ and $c^{\prime}$. The UE sends a measurement report to serving cell $c_0$ at time $m=m_\textrm{0}$ if the preparation condition is fulfilled for the preparation condition monitoring time $T_\textrm{prep}$.

\begin{figure}[!t]
%\vspace{-\baselineskip}
\textit{\centering
\includegraphics[width = 0.97\columnwidth, height = 0.38\columnwidth]{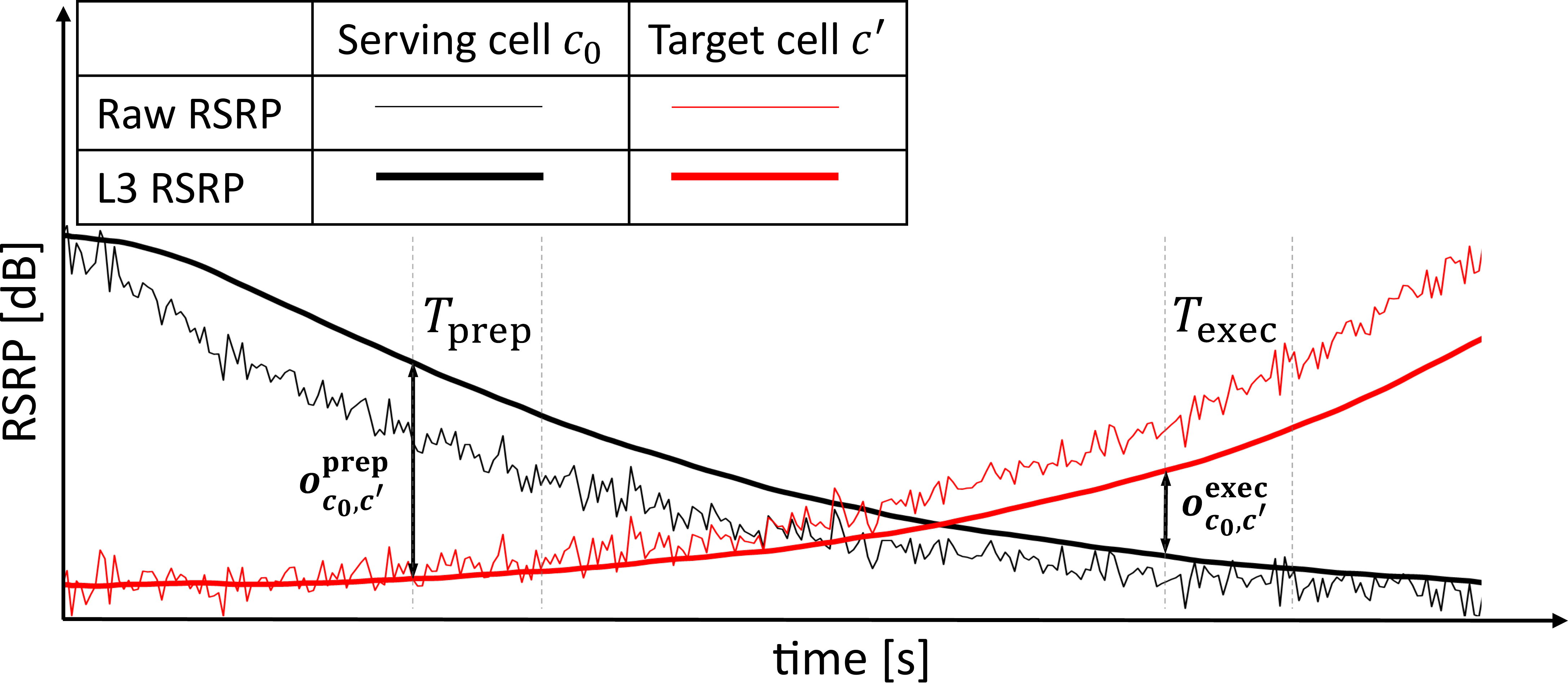}
\vspace{-2pt}
\caption{Illustration of CHO mechanism from serving cell $c_0$ to target cell $c^{\prime}$.} 
\label{fig:Fig0}} 
\vspace{-0.6\baselineskip}
\end{figure}

Once it receives the measurement report, the serving cell initiates the preparation of target cell $c^{\prime}$ over the Xn interface and provides the UE with the conditional configuration of $c^{\prime}$ through an \textit{RRC Reconfiguration} message. As per 3GPP \cite{38331}, the maximum number of prepared cells on the UE side $n_c^{\textrm{max}}$ can be up to eight. The \textit{CHO execution phase} then begins wherein the UE constantly monitors the CHO execution condition, defined as
\begin{equation}
\label{Eq2}
     P_{c_0}^\textrm{L3}(m)  < P_{c^{\prime}}^\textrm{L3}(m) - o^\textrm{exec}_{c_0,c^{\prime}} \ \text{for} \ m_\textrm{1} - T_\textrm{exec} < m < m_\textrm{1},
\end{equation}
where $m_1>m_0$ and $o^\textrm{exec}_{c_0, c^{\prime}}$ is defined as the CHO execution offset between $c_0$ and $c^{\prime}$. The UE executes a handover towards the prepared target cell $c^{\prime}$ if the execution condition at $m=m_\textrm{1}$ is fulfilled for the execution condition monitoring time $T_\textrm{exec}$. An illustration of the CHO mechanism is shown in Fig.\,\ref{fig:Fig0}. 

%%%%%%%%%%%%%%%%%%%%%%
% Here is the first figure.

\subsection{Intra-cell Mobility} \label{Subection2.2}
Intra-cell mobility pertains to a set of L1 and L2 transmit (Tx) beam management procedures to determine and update the serving Tx beam(s) for each UE within a serving cell $c_0$, as defined in \textit{3GPP Release 15} \cite{38802}. The main component is Tx beam selection, where the UE uses network assistance to choose the serving beam $b_0$ that it then uses to communicate with $c_0$ \cite[Sec. II-B]{SubhyalMPUEpaper}. The other key component is beam failure detection, where the UE constantly monitors the radio link quality (RLQ) \cite{38133}, which is determined by further averaging the average downlink SINR using a linear moving-average filter. If the RLQ falls below a certain SINR threshold $\gamma_\textrm{out}$, it is accounted for as a failure of the serving beam. In this case, the UE initiates a beam failure recovery procedure where it tries to recover to another beam of the serving cell. As such, the UE performs a random access attempt on the target beam $b^{\prime}$ that has the highest L1 RSRP beam measurement $P_{c_0,b}^\textrm{L1}(m)$ value and then waits for the BS to send a random access response indicating that the access was successful. In the event of an unsuccessful attempt, the UE tries another random access using $b^{\prime}$. In total, $N_\textrm{BAtt}$ such attempts are made at time intervals of $T_\textrm{BAtt}$. If all attempts fail, a radio link failure (RLF) is declared by the UE.

\subsection{SINR Model} \label{Subection2.3}
The average downlink SINR at the discrete time instant $m$ for Tx beam $b \in B$ of cell $c \in C$ is denoted as $\gamma_{c,b}(m)$. The instantaneous downlink SINR is a random variable and its expected value is referred to as the “average downlink SINR” in \cite{AmanatdownlinkSINR}, for which a closed-form expression is derived. For computational complexity purposes, it is then computed using the Monte-Carlo approximation given for the opportunistic resource-fair scheduler, which maximizes radio resource usage by ensuring that the spare resources of the less crowded beams are fully utilized. As will be seen later in \Cref{Subection3.2} and \Cref{Section4}, the SINR has a key role in our proposed RACH-less handover signaling scheme and the mobility failure models.

%%%%%%%%%%%%%%%%%%%%%%

\section{RACH-less Handover System Model} \label{Section3}

In this section, the handover interruption time is explained in terms of its different components. Thereafter, the proposed RACH-less handover signaling scheme assisted through early TA acquisition is discussed.

\subsection{Handover Interruption Time Breakdown} \label{Subection3.1}
\begin{figure}[!t]
%\vspace{-\baselineskip}
\textit{\centering
\includegraphics[width = 0.94\columnwidth]{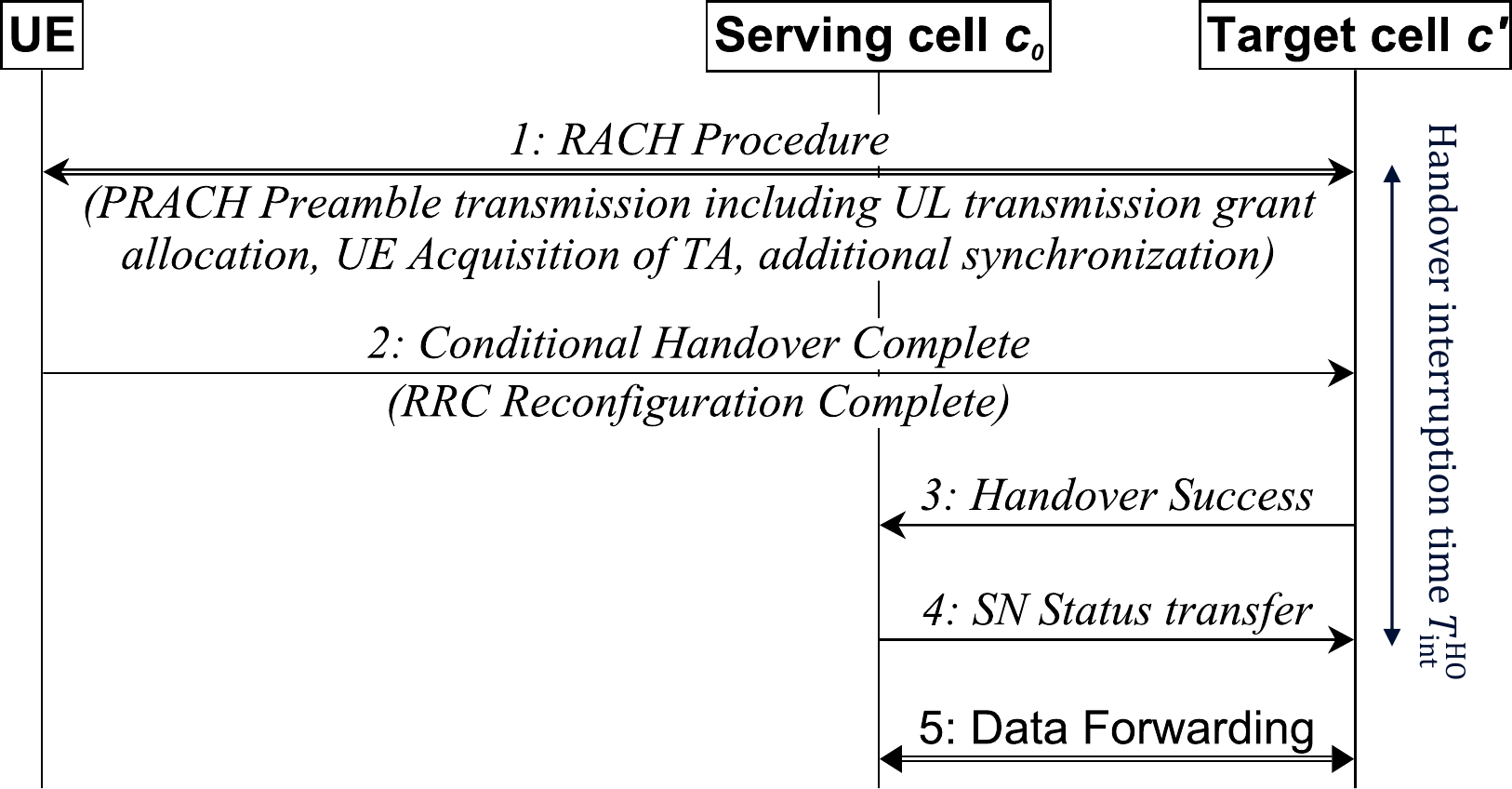}
\vspace{-0.8pt}
\caption{Depiction of the handover interruption time for the CHO mechanism. The control and data plane signaling is indicated in italic and roman font, respectively.} 
\label{fig:Fig0.5}} 
\vspace{-0.6\baselineskip}
\end{figure}

As discussed in \Cref{Section1}, handover interruption time $T^\textrm{HO}_\textrm{int}$ can be described as the outage duration during which the UE is unable to communicate to the network when it is in the process of executing a handover \cite{CHOTechRep}. The handover interruption time regarding signaling exchange is depicted in Fig.\,\ref{fig:Fig0.5}. In line with the recent 3GPP standards \cite[Sec. 9.2.6]{38300}, the RACH procedure is assumed to be 2-step contention-free random access (CFRA). Compared to the 4-step CFRA, 2-step CFRA reduces both the handover interruption time and the signaling overhead. For the CHO mechanism, the handover interruption starts when the CHO execution condition in (\ref{Eq2}) is satisfied. After that, the UE begins detaching from the serving cell $c_0$ and attempting to establish a connection to the target cell $c^{\prime}$. In the first step, the UE sends a preamble associated with the physical random access channel (PRACH) to the target cell to indicate a random access attempt and assist the UE in acquiring the UL transmission grant and the transmit-TA, among other synchronization parameters. This preamble is called the PRACH preamble. The UL transmission grant specifies the time and frequency resources and the associated transport format resources the UE needs to use for the uplink shared channel \cite[Sec. 14.2]{dahlman5Gbook}. The transmit-TA is defined as a UE-specific negative time offset that helps ensure that the uplink transmissions from all UEs in the network are synchronized when received by the BS \cite[Sec. 15.2]{dahlman5Gbook}. If the alignment does not lie within the size of a specific cyclic prefix (CP), the uplink frames would overlap and this would cause intersymbol interference, which may lead to reception failures at the BS. From here on, the term \textit{TA} refers to \textit{transmit-TA}. 

%%%%%%%%%%%%%%%%%%%%%%
% Here is the second figure.

Upon completion of successful random access, the UE sends a \textit{Conditional Handover Complete} message to the target cell $c^{\prime}$. The target cell notifies the serving cell $c_0$ about the accomplished handover procedure through the \textit{Handover Success} message. This is followed by an \textit{Sequence Number (SN) Status Transfer} message from $c_0$ to $c^{\prime}$ that marks the completion of the handover execution process. Both these message exchanges are over the  Xn interface, i.e., the interface between the 5G network entities. Based on the measurements for an experimental LTE network in \cite{NokiaRACHlesspaper}, the delay of \SI{5}{ms} per Xn message is also considered for the 5G network.

For the CHO mechanism, 3GPP \cite{38300} has specified early and late data forwarding, which has an impact on the overall handover interruption time. In early data forwarding, $c_0$ can start forwarding data to $c^{\prime}$ during the \textit{preparation phase}, once it has transmitted the \textit{RRC Reconfiguration message} to the UE, as mentioned in \Cref{Subection2.1}. Herein, the benefit is an interruption time comparable to baseline handover. Whereas, in late data forwarding, the serving cell starts data forwarding only after the reception of the \textit{Handover Success} message from the target cell. The advantage of late data forwarding is that the serving cell only forwards data to a single target cell to which a handover is being executed instead of all multiple prepared target cells that have fulfilled the CHO preparation condition in (\ref{Eq1}). Therefore, we assume late data forwarding in this paper, unless specified otherwise. As such, the handover interruption time lasts till the target cell receives the \textit{SN Status Transfer} message from the serving cell. 

\Cref{Table1} shows a more detailed breakdown of the CHO intra-frequency handover interruption time for FR2, where the total handover interruption time $T^\textrm{HO}_\textrm{int}$ sums up to \SI{54.375}{ms}. It can be seen in \Cref{Table1} that the PRACH preamble transmission associated delay, i.e., components 4, 5, and 6, is \SI{11.375}{ms}. If it is also assumed that the partial synchronization associated with fine time tracking and full timing information of the target can be established beforehand, $T^\textrm{HO}_\textrm{int}$ for the RACH-less case can be reduced from \SI{54.375}{ms} to \SI{40}{ms}. This means that each RACH-less handover has about 25\% less handover interruption time compared to a typical RACH-aided handover.

% %%%%%%%%%%%%%%%%%%%%%%
% \begin{table}[!t]
% \begin{center}
% \caption{\vphantom{\rule[0.0in]{2pt}{\baselineskip}}%
% Components of intra-frequency handover interruption for the CHO mechanism in FR2 \cite{HOInterruptionTimeReductionTechRep}}
%  \vspace{-2.4pt}
% \label{Table1}
% \begin{tabular}{| l | l | l |}
% \hline
% \bfseries \# & \bfseries Description & \bfseries Time (ms) \\
% \hline
% 1 & UE processing time $T_\textrm{processing}$ & 20  \\
% \hline
% 2 & Fine time tracking and acquiring full timing  & 10 \\
% &  information of the target cell $T_\textrm{Delta}$ & (on avg.)  \\
% \hline
% 3 & $T_\textrm{margin}$ (for SSB post-processing) & 2 \\
% \hline
% 4 & Delay in acquiring first available PRACH in target   & 10 \\
%   &  cell $T_\textrm{IU}$ & (on avg.) \\
% \hline
% 5  & PRACH preamble transmission  & 0.125\\
% \hline
% 6  & UL grant allocation and TA for UE  & 1.25 \\
% \hline
% 7 & UE RRC reconfiguration complete message  & 1 \\
% \hline
% 8 & Interruption from late data forwarding (handover  & 10 \\
%   &  success and SN status transfer)   & \\
% \hline
%   &  \textbf{Total handover interruption time $T^\textrm{HO}_\textrm{int}$} & \textbf{55} \\
% \hline
% \end{tabular}
% \end{center}
%  \vspace{-10pt}
% \end{table}
%%%%%%%%%%%%%%%%%%%%%%
\begin{table}[!t]
 \renewcommand{\arraystretch}{0.95}
\begin{center}
\caption{\vphantom{\rule[0.0in]{2pt}{\baselineskip}}%
Components of intra-frequency handover interruption for the CHO mechanism in FR2 \cite{HOInterruptionTimeReductionTechRep}}
 \vspace{-2.4pt}
\label{Table1}
\begin{tabular}{lll}%{| l | l | l |}
% \hline
\toprule
\bfseries \# & \bfseries Description & \bfseries Time (ms) \\
\midrule
%\hline
1 & UE processing time $T_\textrm{processing}$ & 20  \\
%\hline
2 & Fine time tracking and acquiring full timing  & 10 \\
&  information of the target cell $T_{\Delta}$ & (on avg.)  \\
%\hline
3 & $T_\textrm{margin}$ (for SSB post-processing) & 2 \\
%\hline
4 & Delay in acquiring first available PRACH in target   & 10 \\
  &  cell $T_\textrm{IU}$ & (on avg.) \\
%\hline
5  & PRACH preamble transmission  & 0.125\\
%\hline
6  & UL grant allocation and TA for UE  & 1.250 \\
%\hline
7 & UE RRC reconfiguration complete message  & 1 \\
%\hline
8 & Interruption from late data forwarding (handover  & 10 \\
  &  success and SN status transfer)   & \\
%\hline
\midrule
  &  \textbf{Total handover interruption time $T^\textrm{HO}_\textrm{int}$} & \textbf{54.375} \\
\bottomrule
%\hline
\end{tabular}%
\end{center}%
 %\vspace{-10pt}
\end{table}%

\subsection{Proposed Model for Early TA Acquisition} \label{Subection3.2}
\begin{figure}[!t]
%\vspace{-\baselineskip}
\textit{\centering
\includegraphics[width = 0.96\columnwidth]{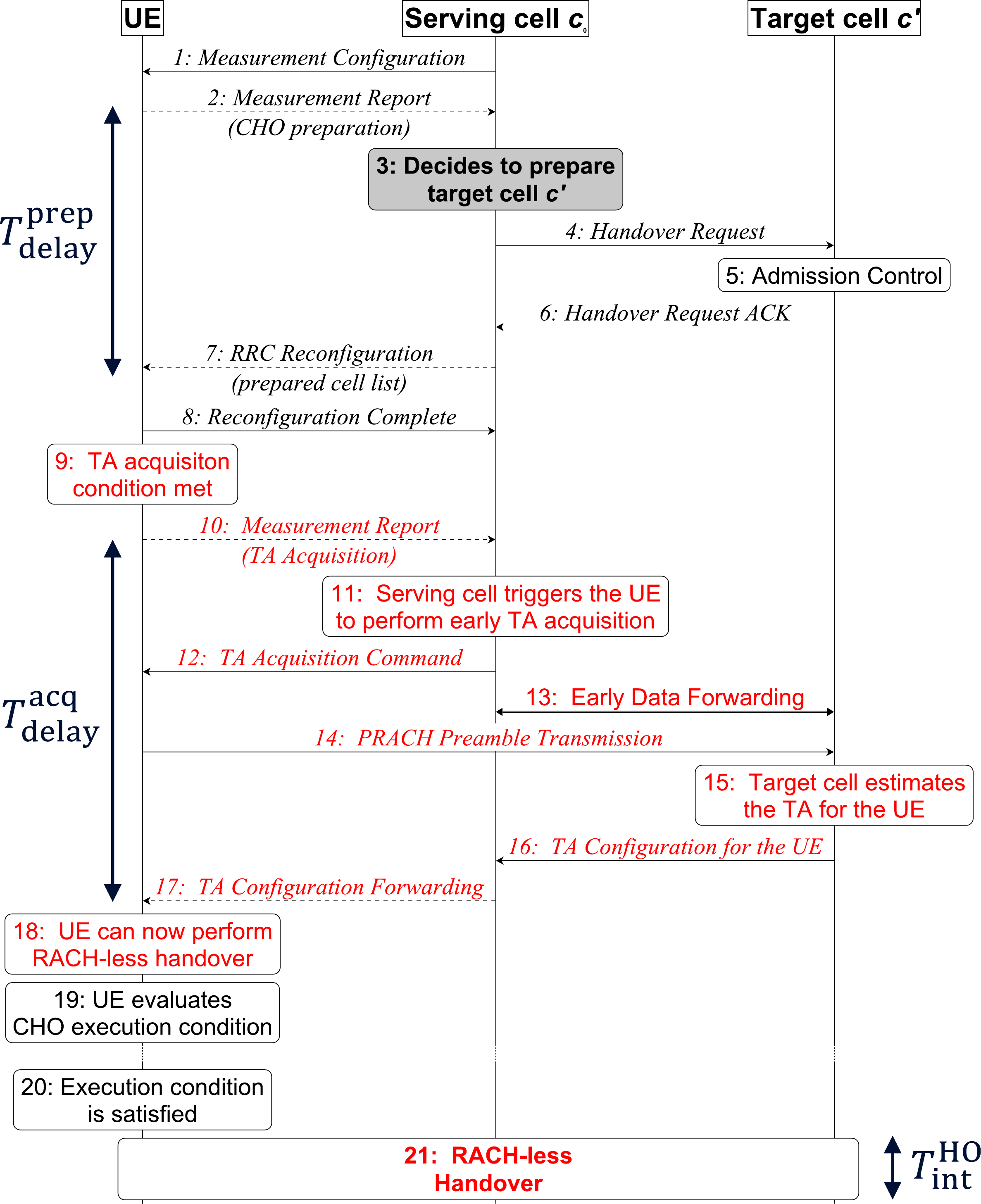}
\vspace{-0.6pt}
\caption{Signaling diagram depicting the proposed enhancement (shown in red) for early acquisition of the TA in CHO to execute a RACH-less handover. The solid and broken lines represent ideal and non-ideal links, respectively.} 
\label{fig:Fig3}} 
\vspace{-1\baselineskip}
\end{figure}

As can be seen in  Fig.\,\ref{fig:Fig0}, there is a significant time delay between the CHO \textit{preparation} and \textit{execution phases} in the decoupled handover mechanism. This delay can be as large as \SI{10}{s} \cite[Sec. I]{JedrzejCHOCFRApaper}. Such a large delay opens the opportunity to commence early TA acquisition  between the CHO \textit{preparation} and \textit{execution phases}. As such, an early TA acquisition model for CHO is formulated where an additional triggering condition for TA acquisition is specified in between the CHO preparation and CHO execution conditions. This signaling diagram for the proposed scheme is illustrated in Fig.\,\ref{fig:Fig3}. \textit{Steps 1} to \textit{7} depict the \textit{CHO preparation phase}. In \textit{Step 4}, however, the serving cell $c_0$ additionally informs the target cell $c^{\prime}$ that the UE would perform early TA acquisition in case the TA acquisition condition is triggered. This would enable $c^{\prime}$ to send the PRACH preamble configuration \cite[Sec. 8]{38213} to $c_0$ in the \textit{Handover Request ACK} in \textit{Step 6}, which it then forwards to the UE in \textit{Step 7}. The PRACH configuration is needed later for the early TA acquisition when the UE sends out the PRACH preamble. The TA acquisition condition is defined as 
\begin{equation}
\label{Eq3}
     P_{c_0}^\textrm{L3}(m)  < P_{c^{\prime}}^\textrm{L3}(m) + o^\textrm{acq}_{c_0,c^{\prime}} \ \text{for} \  m_\textrm{acq} - T_\textrm{acq} < m < m_\textrm{acq},
\end{equation} 
where $o^\textrm{acq}_{c_0,c^{\prime}}$ is defined as the TA acquisition offset between cell $c_0$ and $c^{\prime}$. The UE sends a measurement report to serving cell $c_0$ at time $m=m_\textrm{acq}$ if the TA acquisition condition is fulfilled for the TA acquisition condition monitoring time $T_\textrm{acq}$. This is shown as \textit{Step 10} in Fig.\,\ref{fig:Fig3}. 

%The proposed model assumes that the measurement report can only be successfully delivered to $c_0$ if the average downlink SINR of $c_0$ via the serving beam $b_0$, i.e., $\gamma_{c_0,b_0}(m)$, is above a certain SINR threshold $\gamma_\textrm{out}$. Herein, we assume reciprocity between the uplink and downlink SINR. 

Once the measurement report has been successfully received by the serving cell, it triggers the UE to perform an early TA acquisition. The serving cell sends out a \textit{TA Acquisition Command} to the UE, shown as \textit{Step 12}. %It is assumed that this message also contains the PRACH preamble configuration \cite[Chapter 16]{dahlman5Gbook}. 
In this paper, an enhancement to the existing CHO framework is proposed in \textit{Step 13}, whereby the serving cell $c_0$ can start with early data forwarding \cite{38300} after \textit{Step 12}. As such, the serving cell now selectively forwards data to target cells that have fulfilled the more strict TA acquisition condition in (\ref{Eq3}). In conventional early data forwarding, the serving cell would have executed early data forwarding to all the prepared cells. Besides reducing the signaling overhead, the main benefit of this enhancement comes from the fact that the \SI{10}{ms} interruption from late data forwarding, i.e., component 8 of \Cref{Table1}, can be avoided. Consequently, the handover interruption time $T^\textrm{HO}_\textrm{int}$ reduces further from \SI{40}{ms} to \SI{30}{ms}. 

The UE then proceeds with the \textit{PRACH Preamble Transmission} to the target cell~$c^{\prime}$. Herein, an ideal radio uplink is assumed between the UE and $c^{\prime}$. This is a well-founded assumption because the radio link between the UE and $c^{\prime}$ is sufficient on account of the target cell $c^{\prime}$ having satisfied both the CHO preparation and the more stringent TA acquisition conditions in (\ref{Eq2}) and (\ref{Eq3}), respectively. In line with the MPUE architecture considered in this paper, it is assumed that the UE can use the best panel for target cell $c^{\prime}$ \cite[Sec. III.B] {SubhyalMPUEpaper} to transmit this message. In case the best panel is the same as the serving panel that the UE is using to communicate with the serving cell $c_0$, the UE may use any of the other two panels with the highest L1 beam panel RSRP. 

Since $c^{\prime}$ was already informed by $c_0$ that an early TA acquisition may be attempted by the UE in case the TA acquisition in (\ref{Eq3}) is triggered, the response to \textit{Step 14} is not the conventional \textit{Random Access Response} message from the target cell to the UE \cite{38300}. Instead, once the target cell receives the PRACH preamble, it establishes initial synchronization, including UL transmission grant allocation and an estimation of the TA for the UE, which it then relays to the serving cell in \textit{Step 17}. The serving cell then forwards the TA configuration to the UE in \textit{Step 16}. This two-step relay falls in line with \textit{3GPP Release 18} recommendation \cite{EarlyTAAcqTDoc}. %Herein, the proposed model assumes that the \textit{TA Configuration Forwarding} message can only be delivered successfully to the UE if the serving cell SINR is above $\gamma_\textrm{out}$.

In the proposed model, all signaling exchanges between the UE and serving cell $c_0$ are assumed to be perfect, except for \textit{Step 10} and \textit{Step 17}. Herein, a signaling failure occurs if the average downlink SINR of $c_0$ via the serving beam $b_0$, i.e., $\gamma_{c_0,b_0}(m)$, falls a certain SINR threshold $\gamma_\textrm{out}$, indicating adverse radio conditions. A failure in either of these steps means the UE has failed to acquire the TA. Therefore, it cannot perform a RACH-less handover; instead, it has to perform a typical RACH-aided handover as a fallback measure.

%%%%%%%%%%%%%%%%%%%%%%
% Here is the third figure.

The time alignment timer value is taken as \SI{10.24}{s} as per \cite{38331}. This ensures that the acquired TA remains valid until handover execution, without the need for a TA re-acquisition. As discussed before, this is because the maximum delay between the \textit{CHO preparation} and \textit{execution phases} can be up to \SI{10}{s} but the \textit{TA acquisition phase} starts well after the \textit{CHO preparation} phase. Furthermore, it is assumed that for the mobility scenario considered in this paper, the CP size is chosen large enough such that any uplink timing misalignment does not exceed the CP size \cite[Sec. 15.2]{dahlman5Gbook}. Once the UE has successfully received the TA configuration, it is in a position to perform RACH-less handover. As such, when the CHO execution condition in (\ref{Eq2}) is satisfied, the UE executes a RACH-less handover towards the target cell, shown as \textit{Step 21}. This marks the completion of the handover execution process.

%%%%%%%%%%%%%%%%%%%%%%

\section{Simulation Scenario and Parameters} \label{Section4}
In this section, the simulation setup for the 5G network model is explained with selected simulation parameters that are listed in \Cref{Table2}. A list of the complete simulation parameters can be found in \cite[Sec. IV]{SubhyalMPUEpaper}. The simulations have been performed in our proprietary MATLAB-based system-level simulator \cite{IngoSONtoolpaper}, which elaborately models the handover process using a trigger engine-based setup where multiple handovers can be triggered concurrently.

% \begin{table}[!t]
% \renewcommand{\arraystretch}{1.5}
% \caption{\vphantom{\rule[0.0in]{2pt}{\baselineskip}}Simulation parameters}
% \vspace{-2.6pt}
% \centering
% \begin{tabular}{|l | l | l|}
% \hline
% \bfseries Parameter & \bfseries & \bfseries Value\\
% \hline
% CHO preparation offset & $o^\textrm{prep}_{c_0,c^{\prime}}$  & \SI{5}{dB}\\
% \hline
% CHO execution offset & $o^\textrm{exec}_{c_0,c^{\prime}}$  & \SI{3}{dB}\\
% \hline
% CHO preparation condition monitoring time & $T_\textrm{prep}$ & \SI{80}{ms}\\
% \hline
% CHO execution condition monitoring time & $T_\textrm{exec}$ & \SI{80}{ms}\\
% \hline
% Handover preparation delay & $T^\textrm{prep}_\textrm{delay}$ & \SI{50}{ms}\\
% \hline
% Maximum number of prepared cells & $n_c^{\textrm{max}}$ & 4\\
% \hline
% TA acquisition offset & $o^\textrm{acq}_{c_0,c^{\prime}}$ & \SI{2}{dB} \\
% \hline
% TA acquisition condition monitoring time & $T_\textrm{acq}$  &  \SI{20}{ms} \\
% \hline
% TA acquisition delay & $T^\textrm{acq}_\textrm{delay}$ & \SI{50}{ms}\\
% \hline
% Time alignment timer & $T_{\mathrm{alig}}$ & \SI{10.24}{s}\\
% \hline
% SINR threshold & $\gamma_\textrm{out}$  & \SI{-8}{dB} \\
% \hline
% Time step & $\Delta t$  & \SI{10}{ms}\\
% \hline
% Simulated time & $t_\textrm{sim}$ & \SI{30}{s} \\
% \hline
% \end{tabular}
% \label{Table2}
%  \vspace{-10pt}
% \end{table}
\begin{table}[!t]
\renewcommand{\arraystretch}{1.4}
\caption{\vphantom{\rule[0.0in]{2pt}{\baselineskip}}Simulation parameters}
\vspace{-2.6pt}
\centering
\begin{tabular}{lll}%{|l | l | l|}
%\hline
\toprule
\bfseries Parameter & \bfseries & \bfseries Value\\
\midrule
%\hline
CHO preparation offset & $o^\textrm{prep}_{c_0,c^{\prime}}$  & \SI{5}{dB}\\
%\hline
CHO execution offset & $o^\textrm{exec}_{c_0,c^{\prime}}$  & \SI{3}{dB}\\
%\hline
CHO preparation condition monitoring time & $T_\textrm{prep}$ & \SI{80}{ms}\\
%\hline
CHO execution condition monitoring time & $T_\textrm{exec}$ & \SI{80}{ms}\\
%\hline
Handover preparation delay & $T^\textrm{prep}_\textrm{delay}$ & \SI{50}{ms}\\
%\hline
Maximum number of prepared cells & $n_c^{\textrm{max}}$ & 4\\
%\hline
TA acquisition offset & $o^\textrm{acq}_{c_0,c^{\prime}}$ & \SI{2}{dB} \\
%\hline
TA acquisition condition monitoring time & $T_\textrm{acq}$  &  \SI{20}{ms} \\
%\hline
TA acquisition delay & $T^\textrm{acq}_\textrm{delay}$ & \SI{50}{ms}\\
%\hline
Time alignment timer & $T_{\mathrm{alig}}$ & \SI{10.24}{s}\\
%\hline
SINR threshold & $\gamma_\textrm{out}$  & \SI{-8}{dB} \\
%\hline
Time step & $\Delta t$  & \SI{10}{ms}\\
%\hline
Simulated time & $t_\textrm{sim}$ & \SI{30}{s} \\
%\hline
\bottomrule
\end{tabular}
\label{Table2}
 \vspace{-6pt}
\end{table}

% What parameters to have in the table?
% Mention of MPUE-A3.
%%%%%%%%%%%%%%%%%%%%%%

A network model with an urban-micro cellular deployment consisting of a standard hexagonal grid with seven sites, each divided into three cells, is considered. It is based on \textit{Technical Report 38.901} \cite{38901}, which pertains to channel modeling in FR2. The carrier frequency is 28 GHz, and the inter-site distance is 200 m. %Although the network model investigated in this paper cannot be generalized, it offers an equitable abstraction to existing 5G deployments, as well as beyond-5G future network deployments \cite{38901}. 
It is pertinent to mention that although the hexagonal grid model might differ from realistic deployments, this simplification offers an equitable abstraction to existing 5G as well as beyond-5G networks. Moreover, it allows for comparable results with other studies based on the same standardized model. At the start of the simulation, $N_\textrm{UE}$ = 420 UEs are dropped randomly and follow a 2D uniform distribution over the network, moving at constant velocities along straight lines into random directions \cite[Table 7.8-5]{38901}. %The UE direction is selected randomly at the start. 

\begin{figure}[!t]
%\vspace{-\baselineskip}
\textit{\centering
\includegraphics[width = 0.96\columnwidth]{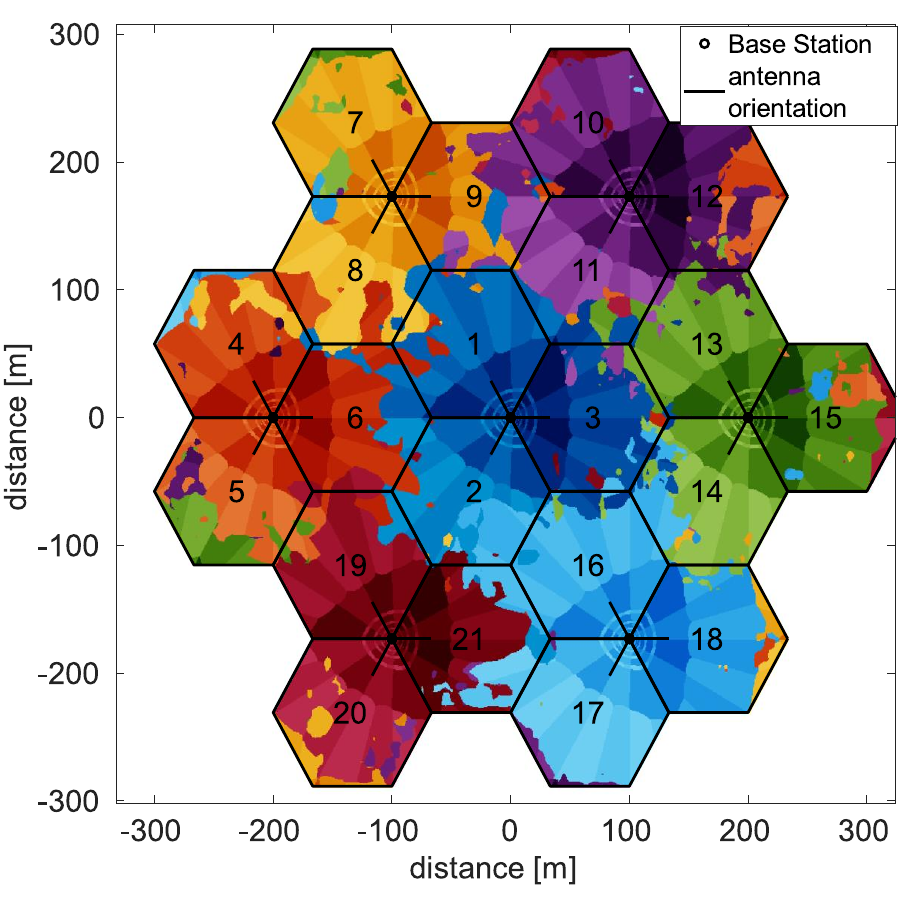}
\vspace{-6pt}
\caption{Simulation scenario consisting of seven hexagonal sites, where each site is serving three cells with 120$^{\circ}$ coverage. The effect of shadow fading is also visible in the form of coverage islands.} 
\label{fig:Figlastadded}} 
\vspace{-1.2\baselineskip} \end{figure}

In line with 3GPP \cite{SubhyalMPUEpaper}, the signal measurement scheme that we consider in this study is MPUE-A3, where it is assumed that the UE can measure the RSRP values from the serving cell $c_0$ and neighboring cells by simultaneously activating all of its three panels. However, the UE only uses only the serving panel for communication with the serving cell \cite[Sec. III.B]{SubhyalMPUEpaper}. The UE speed is 60 km/h, which is the usual speed in the non-residential urban areas of cities~\cite{Rxbeamformingpaper}. % moved it here to the rest of the mobility ;) 
A wrap-around \cite[pp. 140]{IEEE802.16m} is considered, meaning that the hexagonal grid %with seven BS sites 
is repeated around the original %hexagonal grid 
layout in the form of six replicas. The implication is that boundary effects concerning interference are avoided. 
The simulation scenario is shown in Fig.\,\ref{fig:Figlastadded}.

As per 3GPP's study outlined in \textit {3GPP Release 15} \cite{38901}, the channel model takes into account shadow fading as a result of large obstacles and assumes a soft line-of-sight (LoS) for all radio links between the cells and UEs. Soft LoS is defined as a weighted average of the LoS and non-LoS channel components \cite[pp. 59-60]{38901} and is used for both shadow fading and distance-dependent path loss determination. Fast fading is modeled through the low complexity channel model for multi-beam systems proposed in \cite{Umurchannelmodelpaper}, which integrates the spatial and temporal characteristics of 3GPP's geometry-based stochastic channel model \cite{38901} into Jakes’ channel model. The Tx-side beamforming gain model is based on \cite{Umurchannelmodelpaper}, where a 12-beam grid configuration is considered. $K_b=$ 4 beams are simultaneously scheduled for all cells in the network. Beams  $b \in \{9,\ldots,12\}$ have larger beamwidth and relatively smaller beamforming gain and cover regions closer to the BS. Beams $b \in \{1,\ldots,8\}$ have smaller beamwidth and higher beamforming gain and cover regions further apart from the~BS.  
%This can be seen in Fig.\,\ref{fig:Fig4}, where the outer beams are shown in light, and the inner beams are shown in dark color. 

The average downlink SINR $\gamma_{c,b}(m)$ is an indicator of the downlink radio quality and is of key importance in the handover failure and RLF models, which are based on the SINR threshold $\gamma_\textrm{out}$. The former models the failure of a UE to perform a handover from the serving cell $c_0$ to the target cell $c^{\prime}$ and the latter models the failure of a UE while it is still in $c_0$. They are explained in greater detail in \cite[Sec. IV]{SubhyalMPUEpaper}. In 3GPP terminology\cite{38133}, $\gamma_\textrm{out}$, denoted as $\mathrm{Q_{out}}$, is defined as the level at which the radio link can not be reliability received by the UE in the downlink. It corresponds to an out-of-sync block error rate of 10\% and is derived based on hypothetical physical downlink control channel (PDCCH) transmission parameters listed in \cite[Table 8.1.2.1-1]{38133}. Through our extensive stimulative investigations centered on the FR2 simulations in the given network model, the threshold $\gamma_\textrm{out}$ has been determined as \SI{-8}{dB}.

%%%%%%%%%%%%%%%%%%%%%%

\section{Performance Evaluation} \label{Section5}
In this section, the mobility performance of the proposed RACH-less handover signaling scheme is compared against the typical RACH-aided handover for the CHO mechanism. The key performance indicators (KPIs) used for comparison are explained below.

\subsection{KPIs} \label{Subection5.1}

\textit{Total successful handovers}: Sum of the total number of successful handovers from the serving to the target cells in the network. This is the sum of \textit{RACH-less handovers} and \textit{RACH-aided handovers} in the network.

\textit{Mobility failures}: Sum of the total number of handover failures and RLFs in the network.

\textit{Radio / Xn interface signaling overhead}: The total number of signaling messages over the radio / Xn interface.

%\textit{Xn interface signaling overhead}: The total number of signaling messages over the Xn interface.

The KPIs above are normalized to number of UEs $N_\mathrm{UE}$ in the network and to time and, thus, expressed as KPI/UE/min.

\textit{Outage:} Outage is defined as the time period when a UE is unable to communicate with the network. Simulation-based investigations based in FR2 networks with the mobility scenario that we considered have shown that the most common type of outage is \textit{outage due to handover interruption} \cite[Sec. V.B]{Rxbeamformingpaper}, which occurs when a UE is in the process of performing a handover. Besides, if the handover failure timer $T_{\mathrm{HOF}}$ expires and the UE declares a handover failure or the RLF timer $T_{\mathrm{RLF}}$ expires and the UE declares an RLF, the UE initiates connection re-establishment \cite{SubhyalMPUEpaper} and this is also accounted for as outage. Hereby, the UE may re-connect to the previous serving cell or another neighboring cell. Another type of outage modeled is when the average downlink SINR of the serving cell $\gamma_{c_0, b_0}$ falls below $\gamma_\mathrm{out}$. Here, it is assumed that the UE cannot communicate with the network, given the adverse radio conditions. The last two types of outages are classified to as \textit{outage due to other mobility events}. Outage is denoted in terms of a percentage 
\vspace{-1mm}
\begin{equation}
\label{Eq6} 
\textrm{Outage} \ (\%) = \frac{\sum_{u}{\textrm{Outage duration of UE}} \ u} {N_\mathrm{UE} \ \cdot \ \textrm{Simulated time}} \ \cdot \ 100. 
\end{equation}

\subsection{Simulation Results} \label{Subection5.2}
\begin{figure}[!t]
\textit{\centering
\includegraphics[width = 0.96\columnwidth]{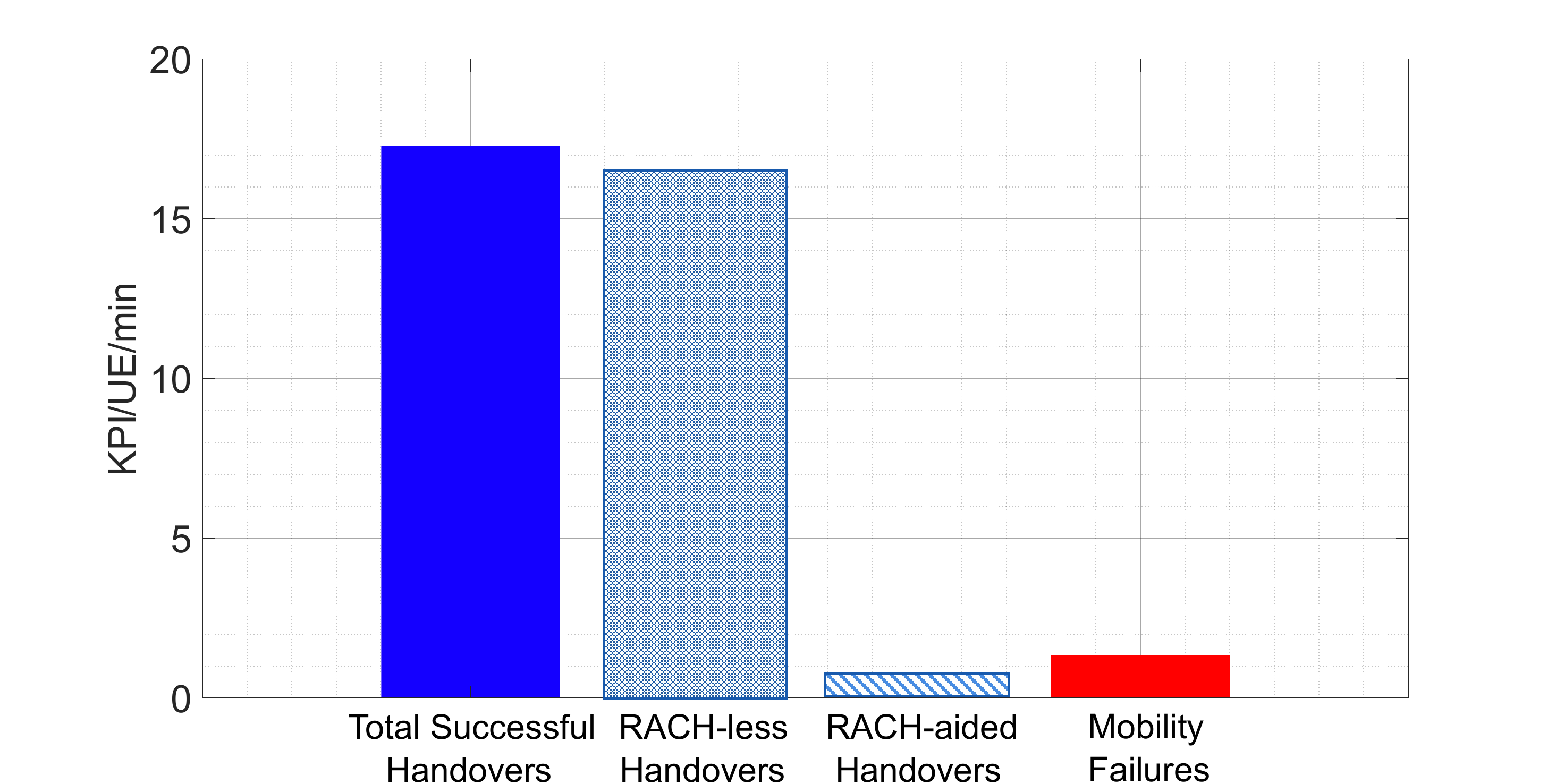}
%\vspace{-0.2\baselineskip}  
\caption{A depiction of the mobility performance of the proposed RACH-less handover scheme.} 
\label{fig:Fig5}}  \vspace{-0.6\baselineskip} 
\end{figure}

Fig.\,\ref{fig:Fig5} shows a mobility performance analysis of the proposed RACH-less handover model for CHO. The total number of successful handovers is 17.3 KPI/UE/min. Out of this, almost 95\% of the handovers are RACH-less on account of the UE having successfully acquired an early TA before being able to execute the handover. The 5\% typical RACH-aided handovers stem from either the failure of the UE to send out a measurement report to the serving cell $c_0$ or a failure of $c_0$ to successfully deliver the TA configuration to the UE, both due to adverse radio conditions whereby the UE then resorts to a RACH-aided handover as a fallback measure. These are shown as \textit{Step 10} and \textit{Step 16} in Fig.\,\ref{fig:Fig3}. The reason for the significantly low failure rate is that the TA acquisition process lies nicely in between the CHO \textit{preparation} and \textit{execution phases}, where the radio link between the UE and the serving cell is sufficiently good on most occasions. As such, the results also show the efficacy of the scheme and the chosen parameters in guaranteeing that most handovers will be RACH-less. The total mobility failures are around 8\% compared to the total number of successful handovers, and this relatively low number of failures is one of the well-known benefits of the decoupled CHO process \cite{JedrzejCHOCFRApaper}.

%%%%%%%%%%%%%%%%%%%%%%
% Here is the fifth figure.

\begin{figure}[!t]
\textit{\centering
\includegraphics[width = 0.96\columnwidth]{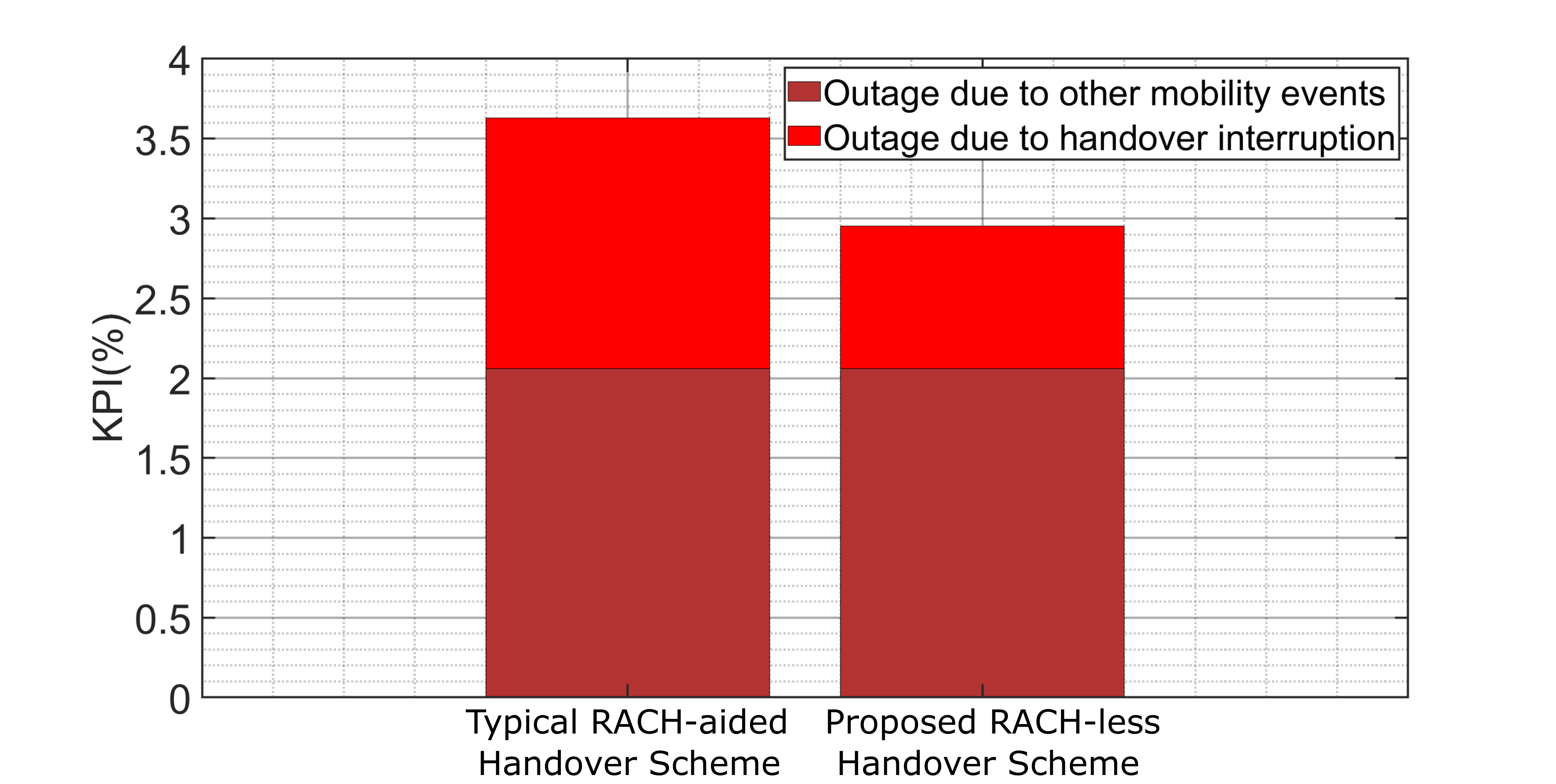}
\vspace{-0.1\baselineskip}  
\caption{A comparison of the outage between the proposed RACH-less handover scheme with the RACH-aided handover for the CHO mechanism. The two different outages shown here add up as the total outage.} 
\label{fig:Fig6}} 
\vspace{-0.2\baselineskip} \end{figure}

The outage comparison between the proposed scheme and the typical RACH-aided handover for the CHO mechanism is shown in Fig.\,\ref{fig:Fig6}. For the typical scheme, it is observed that on average a UE remains in outage for 3.6\% of its $t_\textrm{sim}$ = \SI{30}{s} motion duration. When the proposed scheme is compared with the typical RACH-aided handover scheme, it is seen that there is a 43.2\% relative reduction in the outage due to handover interruption. As discussed in \Cref{Subection3.2}, this is because in the RACH-less case the intra-frequency handover interruption time for CHO in FR2 reduces from $T^\textrm{HO}_\textrm{int}$ = \SI{54.375}{ms} to $T^\textrm{HO}_\textrm{int}$ = \SI{30}{ms}. The reduction in the handover interruption time is reflected as an 18.7\% reduction in the total outage.

\begin{figure}[!t]
\textit{\centering
\includegraphics[width = 0.96\columnwidth]{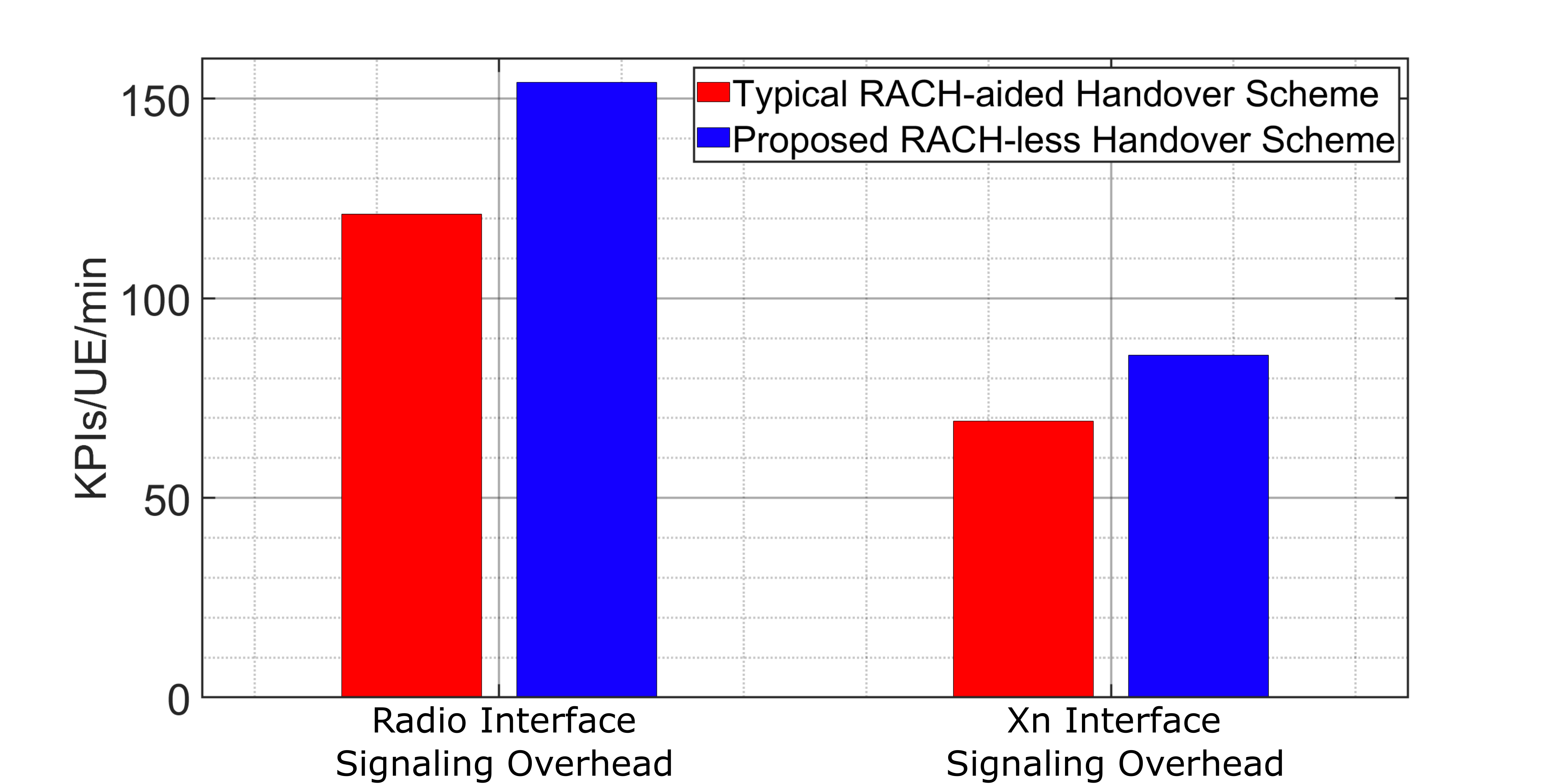}
%\vspace{-0.2\baselineskip}  
\caption{A comparison of the signaling overhead due to the handover process between the proposed RACH-less handover scheme with the RACH-aided handover for the CHO mechanism.} 
\label{fig:Fig7}} 
\vspace{-0.6\baselineskip} \end{figure}

As with most enhancements, the proposed scheme has some drawbacks. As shown in Fig.\,\ref{fig:Fig3}, the proposed enhancement to acquire an early TA of the target cell involves signaling messages (shown in red) over both the radio and Xn interfaces in the 5G network. This signaling is on top of the existing signaling needed for the CHO \textit{preparation} and \textit{execution phases}. It is pertinent to mention here that for the RACH-aided handovers, the 2-step CFRA scheme \cite[Sec. 9.2.6]{38300} is more signaling-intensive than the RACH-less handovers. Here, the RACH-aided handovers are part of both the typical RACH-aided handover scheme as well as for the proposed RACH-less handover scheme, when the UE has to fallback to a RACH-aided handover. The signaling overhead is quantified in Fig.\,\ref{fig:Fig7}, where it is seen that with the proposed RACH-less handover scheme the signaling overhead over the radio and Xn interfaces increases by about 27\% and 20\%, respectively, when compared to the typical RACH-aided handover.

%%%%%%%%%%%%%%%%%%%%%%

% Here is the sixth figure. 

%%%%%%%%%%%%%%%%%%%%%%

% Here is the seventh figure.

\section{Conclusion} \label{Section6}
    In this paper, a random access channel (RACH)-less handover signaling scheme based on early acquisition of the timing advance (TA) of the target cell is proposed for the conditional handover mechanism. The proposed scheme largely curtails the RACH process and this paper analyses its mobility performance in terms of reducing the handover interruption time and, therefore, the total outage. In the context of mobility studies based on 5G-beamformed networks, such studies are essential because outage is often analyzed solely through the lens of mobility failures. However, our previous studies \cite{Rxbeamformingpaper} have shown that for the given mobility scenario, outage is indeed dominated by the handover process itself which is needed to ensure seamless network connectivity and cannot be done away with. The proposed enhancement exploits the split within the conditional handover mechanism to devise a signaling scheme whereby the user equipment (UE) can establish initial synchronization and acquire the TA of the target cell via the serving cell when both the UE-serving cell link and the UE-target cell links are sufficiently strong. Results have shown that with the proposed RACH-less handover scheme, the UE is mainly successful in the early acquisition of the TA. As a consequence, the outage and its constituent handover interruption time are reduced relatively by 18.7\% and 43.2\%, respectively. This improvement comes at a downside of a nominal increase in the signaling overhead, which is also investigated in the paper and is one topic open to future studies. TA validation based on a comparison of the difference in pathloss values with a threshold is also an interesting research direction \cite[Sec. 5.7.17]{38331}. Another topic open to future studies is the physical random access channel (PRACH) preamble detection at the target cell based on \textit{3GPP Release 18}\cite[Sec. 8.4]{38141-2}, whereby a model is developed for the PRACH preamble transmission that is needed to aid an early acquisition of the TA for a RACH-less handover to occur. 

%\vspace{-0.5\baselineskip} 
\bibliographystyle{IEEEtran}
\bibliography{references}  

\end{document}